\documentclass[envcountsame]{llncs}

\usepackage{amsmath}
\usepackage{latexsym}
\usepackage{theorem}
\usepackage{amssymb}
\usepackage{url}

\newcommand{\comment}[1]{}
\newcommand{\bu}{$\bullet$}
\newcommand{\hs}[1][3ex]{\hspace*{#1}}
\newcommand{\vs}[1][1mm]{\vspace*{#1}}
\newcommand{\moins}{\setminus}

\renewcommand{\a}{\rightarrow}

\renewcommand{\to}{\mapsto}

\newcommand{\ex}{\exists}
\newcommand{\all}{\forall}

\newcommand{\sle}{\subseteq}

\newenvironment{rew}[1][\a]%
  {$\begin{array}{r@{~~#1~~}l}}%
  {\end{array}$}
  {\begin{center}\begin{rew}[#1]}%
  {\end{rew}\end{center}}

{\theorembodyfont{\rmfamily} 

}

\newenvironment{lstgeneric}[2]
  {\begin{list}{#1}{\topsep=.5mm\itemsep=.5mm\parsep=0mm%
    \itemindent=-3ex\labelsep=1ex\labelwidth=0ex #2}}
  {\end{list}}

\newenvironment{lst}[1]
  {\begin{lstgeneric}{#1}{\itemindent=-1ex}}
  {\end{lstgeneric}}

\newenvironment{isa}{\vs[1mm]\small\tt}{\vs[1mm]}
\renewcommand{\i}[1]{{\small\tt #1}}
\newcommand{\pair}[1]{$\{\!|$#1$|\!\}$}
\newcommand{\In}{$\in$ }
\newcommand{\Notin}{$\notin$ }
\newcommand{\arr}{$\Rightarrow$ }
\newcommand{\imp}{$\longrightarrow$ }
\newcommand{\Imp}{$\Longrightarrow$ }
\newcommand{\Hyp}[1]{$[\![$#1$]\!]$}
\newcommand{\Ex}{$\ex$ }
\newcommand{\All}{$\all$ }

\begin{document}


\title{\bf An Isabelle formalization of\\ protocol-independent secrecy\\
with an application to e-commerce}

\author{Fr\'ed\'eric Blanqui}

\institute{Laboratoire d'Informatique de l'\'Ecole Polytechnique\\
91128 Palaiseau Cedex, France\\
\url{http://www.lix.polytechnique.fr/~blanqui/}}

\maketitle

\begin{abstract}
A protocol-independent secrecy theorem is established and applied to
several non-trivial protocols. In particular, it is applied to
protocols proposed for protecting the computation results of
free-roaming mobile agents doing comparison shopping. All the results
presented here have been formally proved in Isabelle by building on
Larry Paulson's inductive approach. This therefore provides a library
of general theorems that can be applied to other protocols.
\end{abstract}


\section{Introduction}

Cryptographic protocols are intended to ensure properties like
secrecy, authentication, anonymity, integrity or
non-repudiability. Initially proposed for securing communications,
they have been more recently proposed for protecting the computation
results of free-roaming mobile agents doing comparison shopping
\cite{yee98mos}. Experience shows that even simple protocols are
difficult to set correctly \cite{lowe95ipl}. Their correctness clearly
needs to be checked by mechanical tools. But this cannot be an easy
task. For instance, secrecy has been shown undecidable even under very
weak assumptions \cite{durgin99fmsp}. And correctness is always with
respect to a given model. A protocol correct in a model may be
incorrect in a richer model \cite{paulson01jcs}. As for the
effectiveness of cryptography, which is often assumed perfect in the
formal versions of protocols, fortunately, it may be precisely related
to the one of real protocols \cite{guttman01ccs}. Although we cannot
expect formal methods to be able to deal with every possible problem
(like \cite{song01uss} for instance), yet they provide better
confidence and may serve in finding protocol-specific flows.

Many different approaches have been proposed so far. Approaches based
on model-checking are quite effective but the state explosion forces
the model to be kept simple, typically by limiting the number of
agents and other parameters to one or two although, in some cases,
checking can be done in a finite model \cite{lowe98csf}. Approaches
based on proof assistants allow a finer and more general modelization
but require much effort, typically of several days or weeks, although
the use of automated theorem provers may be very effective too
\cite{cohen00csf}. Proof-assistants (Isabelle \cite{nipkow02lncs}, PVS
\cite{shankar96fmcad}, etc.) can establish protocol-independent
theorems whose conditions could be proved by means of specialized
automatic tools, providing a really complete certification. In this
spirit, Millen and Rue{\ss} formalized a protocol-independent secrecy
theorem \cite{millen00ssp} in PVS \cite{millen00fmcs} and later
developed with Cortier an automatic proof search procedure
\cite{cortier01csf}.

In this paper, we develop in Isabelle a more general secrecy theorem
based on Paulson's inductive approach \cite{paulson98jcs}, which has
several advantages over the previous approach. First, we require no
additional information ({\em spell} or {\em state events}) in order to
be more general and also compatible with the libraries already
implemented in Isabelle by Paulson and Bella. Second, we establish a
first secrecy theorem (Section \ref{sec-guard}) which only involves
the Isabelle theory of messages, and thus is independent of any
protocol formalization. It is based on the simple and intuitive notion
of {\em guarded messages} which, as opposed to the notion of {\em
coideal} used in \cite{millen00ssp}, distinguishes between what must
be kept secret (a key or a nonce \i{n}) from how this secret is
ensured (the set \i{Ks} of the keys whose inverses are used for
encrypting \i{n}). It simply says that, if someone can get \i{n} by
decomposing and decrypting a set of guarded messages, then he must
also be able to get one of the key of \i{Ks}. Therefore, if the keys
of \i{Ks} are not compromised, then the spy cannot get \i{n}.

Then, we establish a second theorem (Section \ref{sec-proto}) showing
that, for proving the guardedness of a nonce or a key \i{n} used in a
protocol, and hence that \i{n} is kept secret, it is sufficient to
prove that each rule of the protocol indeed preserve the guardedness
of \i{n}, without having to care about what the spy can do (which is
taken into account once and for all in the proof of the theorem). This
departs from proofs in strand spaces \cite{guttman98ssp} or with
coideals \cite{cortier01csf} where it is necessary to step back into
the protocol rules, including the rules for the spy, in order to
explore every possibility of how certain message fields could have
been published.

We applied our results to several well-known protocols:
Needham-Schoeder-Lowe \cite{needham78,lowe95ipl}, Otway-Rees
\cite{otway87} and Yahalom \cite{abadi90rs}. We also verified two
protocols proposed by Asokan, G\"ulc\"u and Karjoth \cite{asokan98ma}
for protecting the computation results of free-roaming mobile agents
doing comparison shopping. These have never been formally certified
before. The properties claimed by the authors (see Section
\ref{sec-def}) not only include confidentiality properties but also
non-repudiability and integrity properties. This shows a new
application of Isabelle.

The paper is organized as follows. In Section \ref{sec-induc}, we
quickly present Paulson's inductive approach \cite{paulson98jcs}. In
Section \ref{sec-guard}, we present our notion of {\em guarded
message} for proving secrecy. In Section \ref{sec-proto}, we introduce
a more precise formalization of protocols and show how it helps to
prove guardedness. In Section \ref{sec-def}, we present the protocols
P1 and P2 proposed in \cite{asokan98ma} for protecting the computation
results of free-roaming agents. In Section \ref{sec-proof}, we explain
the formal proof of their correctness.

All the Isabelle files are freely available on our web page.


\section{Paulson's inductive approach}
\label{sec-induc}

The inductive approach has been introduced by Paulson
\cite{paulson98jcs} and applied to many non-trivial protocols within
the generic proof assistant Isabelle \cite{nipkow02lncs}: Kerberos IV
\cite{bella98esorics}, TLS \cite{paulson99css}, SET
\cite{paulson01ijcar}, etc. All these results are part of the Isabelle
distribution and can be freely used to certify other protocols.

In this approach, a protocol is represented as the set of all the
possible sequences of events that can occur by following the protocol
steps and by having a spy able to send fake messages built from the
analysis of past traffic. An infinite number of agents is assumed:

\begin{isa}
\noindent datatype agent = Server | Friend nat | Spy
\end{isa}

Messages are represented as the elements of the following inductive
data type:

\begin{isa}
\noindent datatype msg = Number nat (* guessable *)
| Nonce nat (* not guessable *)\\
| Agent agent | Key nat | Hash msg | \pair{msg, msg} | Crypt key msg
\end{isa}

This has several important consequences:

\begin{lst}{\bu}
\item Encryption is assumed to be perfect: decryption can only be done
by using the inverse of the key used for encryption. Within a
public-key infrastructure, the inverse of the public key of an agent
\i{A}, \i{pubK A}, is its private key \i{priK A}. Otherwise, each
agent \i{A} is assumed to have a symmetric key \i{shrK A} whose
inverse is itself.

\item Hashing is collision-free: two distinct messages give two distinct
hash codes. This comes from the fact that the constructors of an
inductive data type are injective.

\item Each kind of message is recognizable and hence cannot be
confused with another one: an agent name \i{Agent A} is distinct from
a key \i{Key K} or an encrypted message \i{Crypt K X}, etc. This comes
from the fact that the constructors of an inductive data type have
distinct images. This implies that encryption schemes for which there
are relations between encrypted messages like XOR (\i{Crypt K (Crypt K
X) = X}) or RSA (\i{Crypt K (Crypt K' X) = Crypt K' (Crypt K X)})
cannot be formalized.
\end{lst}

Then, Paulson introduces three important functions:

\begin{lst}{\bu}
\item The function \i{parts H} returns the set of all the actual
sub-components of a set \i{H} of messages (\i{X} is not a
sub-component of \i{Hash X}):

\begin{isa}
consts parts :: "msg set \arr msg set"\\
inductive "parts H" intros\\
Inj: "X \In H \Imp X \In parts H"\\
Fst: "\pair{X,Y} \In parts H \Imp X \In parts H"\\
Snd: "\pair{X,Y} \In parts H \Imp Y \In parts H"\\
Body: "Crypt K X \In parts H \Imp X \In parts H"
\end{isa}

\item The function \i{analz H} returns the set of all the sub-components
that can be obtained by decomposing and decrypting (if the necessary
key has itself been obtained) all the messages of the set \i{H}:

\begin{isa}
consts analz :: "msg set => msg set"\\
inductive "analz H" intros\\
Inj: "X \In H \Imp X \In analz H"\\
Fst: "\pair{X,Y} \In analz H \Imp X \In analz H"\\
Snd: "\pair{X,Y} \In analz H \Imp Y \In analz H"\\
Decrypt: "\Hyp{Crypt K X \In analz H; Key(invKey K) \In analz H}\\\hs
\Imp X \In analz H"
\end{isa}

\item The function \i{synth} returns the set of all the messages that
can be synthesized from a set \i{H} of messages (new nonces and new
keys cannot be synthesized):

\begin{isa}
consts synth :: "msg set \arr msg set"\\
inductive "synth H" intros\\
Inj: "X \In H \Imp X \In synth H"\\
Agent: "Agent agt \In synth H"\\
Number: "Number n \In synth H"\\
Hash: "X \In synth H \Imp Hash X \In synth H"\\
Pair: "\Hyp{X \In synth H; Y \In synth H} \Imp \pair{X,Y} \In synth H"\\
Crypt: "\Hyp{X \In synth H; Key K \In H} \Imp Crypt K X \In synth H"
\end{isa}
\end{lst}

Finally, the fact that an agent \i{A} sends to another agent \i{B} a
message \i{X} is represented by the event \i{Says A B X}.

As an example, we give the formalization of the Needham-Schroeder-Lowe
protocol \cite{lowe95ipl}:

\begin{lst}{}
\item 1. $A\a B: \{Na,A\}_{Kb}$
\item 2. $B\a A: \{Na,Nb,B\}_{Ka}$
\item 3. $A\a B: \{Nb\}_{Kb}$
\end{lst}

\begin{isa}
\noindent consts nsl :: "event list set"\\
inductive nsl intros\\
Nil: "[] \In nsl"\\
Fake: "\Hyp{evs \In nsl; X \In synth(analz(spies evs))}\\\hs
  \Imp Says Spy B X \# evs \In nsl"\\
NS1: "\Hyp{evs \In nsl; Nonce NA \Notin used evs}\\\hs
  \Imp Says A B (Crypt (pubK B) \pair{Nonce NA, Agent A}) \# evs \In nsl"\\
NS2: "\Hyp{evs \In nsl; Nonce NB \Notin used evs;\\\hs
  Says A' B (Crypt (pubK B) \pair{Nonce NA, Agent A}) \In set evs}\\\hs
  \Imp Says B A (Crypt (pubK A) \pair{Nonce NA,Nonce NB,Agent B})
  \# evs \In nsl"\\
NS3: "\Hyp{evs\,\In nsl; Says A B (Crypt (pubK B)
  \pair{Nonce NA,Agent A}) \In set evs;\\\hs
  Says B' A (Crypt (pubK A) \pair{Nonce NA, Nonce NB, Agent B})
  \In set evs}\\\hs
  \Imp Says A B (Crypt (pubK B) (Nonce NB)) \# evs \In nsl"
\end{isa}

The first two rules are in any protocol. \i{Nil} means that the empty
trace \i{[]} is in the protocol. \i{Fake} means that the spy can send
fake messages built from the analysis of past traffic. The next rules
are the rules of the protocol. \i{set evs} is the set of all the
messages sent so far. \i{used evs} denotes the parts of all the
messages sent so far. \i{ev \# evs} is the list of head \i{ev} and
tail \i{evs}.

Then, one may prove for example that, if \i{B} sends the message of
step 2 and receives the message of step 3, and if the private keys of
\i{A} and \i{B} are not compromised (\i{A} and \i{B} do not belong to
the set \i{bad} of bad agents), then the protocol must have been
initiated by \i{A}:

\begin{isa}
\noindent lemma "\Hyp{A \Notin bad; B \Notin bad; evs \In nsl;\\
Says B A (Crypt (pubK A) \pair{Nonce NA, Nonce NB, Agent B}) \In set evs;\\
Says A' B (Crypt(pubK B) (Nonce NB)) \In set evs}\\
\Imp Says A B (Crypt (pubK B) \pair{Nonce NA, Agent A}) \In set evs"
\end{isa}

As most authentication properties, it relies on the fact that the
nonces used by the agents and encrypted with their public keys are not
known by the spy:

\begin{isa}
\noindent lemma "\Hyp{A \Notin bad; B \Notin bad; evs \In nsl;\\
Says A B (Crypt (pubK B) \pair{Nonce NA, Agent A}) \In  set evs}\\
\Imp Nonce NA \Notin analz(spies evs)"
\end{isa}

This is why, in the following, we concentrate on secrecy proofs.


\section{Guarded messages}
\label{sec-guard}

The idea of the protocol-independent secrecy theorem is simple and
intuitive. In any protocol, secrecy is ensured by using cryptography
and by making sure that the key used for the encryption is indeed
safe. We simply turn this into a formal definition, the notion of {\em
guarded message}, and formally prove that getting the secret indeed
requires to know the key used for encryption (normally safe).


\begin{definition}[Guarded messages]
A nonce or a key \i{n} is {\em guarded} in a message \i{X} by a set of
keys \i{Ks} if every occurrence of \i{n} in \i{X} is inside a
sub-message encrypted by the inverse of a key of \i{Ks}. We denote by
\i{guard n Ks} the set of messages in which \i{n} is guarded by
\i{Ks}.

\begin{isa}
\noindent consts guard :: "nat \arr key set \arr msg set"\\
inductive "guard n Ks" intros\\
No\_Nonce: "Nonce n \Notin parts {X} \Imp X \In guard n Ks"\\
Guard: "invKey K \In Ks \Imp Crypt K X \In guard n Ks"\\
Crypt: "X \In guard n Ks \Imp Crypt K X \In guard n Ks"\\
Pair: "\Hyp{X \In guard n Ks; Y \In guard n Ks}
  \Imp \pair{X,Y} \In guard n Ks"
\end{isa}
\end{definition}


The secrecy theorem can then be stated as follows:

\begin{theorem}[Secrecy]
Let \i{G} be a set of messages where \i{Nonce n} is guarded by a set
of keys \i{Ks}. If \i{Nonce n} belongs to \i{analz G} then there
exists a key \i{K} in \i{Ks} that belongs to \i{analz G}:

\begin{isa}
\noindent theorem Guard\_invKey: "\Hyp{Nonce n \In analz G; Guard n Ks G}\\
\Imp \Ex K. K \In Ks \& Key K \In analz G"
\end{isa}
\end{theorem}

\begin{proof}
Although this result may seem obvious, it is not straightforward to
prove in Isabelle or in any other proof assistant. We explain how our
formal proof proceeds.

First, only a finite part of \i{G} needs to be analyzed to get
\i{Nonce n}. Therefore, we can assume that \i{G} is finite. We then
associate to any finite set \i{G} a measure $\mu($\i{G}$)$ which is
the number of encrypted messages in \i{G}. We can then reason by
induction on $\mu($\i{G}$)$.

Second, to help reason about \i{analz}, we proved the following very
useful decomposition theorem: \i{analz G} = \i{pparts G} $\cup$
\i{analz} (\i{kparts G}), where \i{pparts G} is all the pairs of
\i{G}, their components that are themselves pairs and so on, and
\i{kparts G} is all the components that are not pairs.

Now, since all the messages in \i{G} are guarded, to get \i{Nonce n},
there must be an encrypted message \i{Crypt K Y} in \i{kparts G} such
that \i{Key} (\i{invKey K}) belongs to \i{kparts G} also: we must
decrypt at least one encrypted message for reaching \i{Nonce n}. If
\i{K} is a key of \i{Ks} then we are done. Otherwise, let \i{H} =
\i{kparts G} $\moins$ $\{$\i{Crypt K Y}$\}$. Then, by definition of
\i{analz}, \i{Nonce n} must belong to \i{analz} (\i{H} $\cup$
$\{$\i{Y}$\}$) whose measure is strictly smaller than the one of
\i{G}. Therefore, we can conclude by induction hypothesis.\qed
\end{proof}


Our notion of guardedness has several advantages over the notion of
coideal of Millen and Rue\ss{} \cite{millen00ssp}. First, it is
defined inductively while a coideal is defined as the complement of an
ideal \cite{guttman98csf} (which is defined inductively). Second, it
is more general in the sense that the coideal of $\{$\i{Nonce n}$\}$
$\cup$ \i{Ks} is included in \i{guard n Ks} but not the converse. For
instance, while \i{Key K} is guarded by $\{$\i{K}$\}$ in \i{Crypt}
(\i{invKey K}) (\i{Key K}), it does not belong to the coideal of
$\{$\i{K}$\}$. This is due to the fact that there is no separation
between what must be kept secret (\i{n} in \i{guard n Ks}) and how it
must be kept secret (\i{Ks} in \i{guard n Ks}). Third, guardedness
enjoys a nice monotonicity property:

\begin{isa}
\noindent
lemma guard\_extend: "Ks $\sle$ Ks' \Imp guard n Ks $\sle$ guard n Ks'"
\end{isa}

\noindent
which is very useful in the case of protocols like Yahalom where a
nonce must be guarded by a session key issued by a server (see
below). This is not the case with coideals. For instance,
\pair{\i{Crypt} (\i{pubK~A}) (\i{Nonce~n}), \i{Key~K}}, in which
\i{Nonce~n} is guarded by $\{$\i{Nonce~n, priK~A}$\}$, hence by
$\{$\i{Nonce~n, priK~A, Key~K}$\}$ too, belongs to the coideal of
$\{$\i{Nonce~n, priK~A}$\}$ but not to the coideal of $\{$\i{Nonce~n,
priK~A, Key~K}$\}$. So, guardedness is a finer and more intuitive
notion than the one of coideals.


In the case of the Needham-Schroeder-Lowe protocol, one can formally
prove that \i{NA} and \i{NB} are both guarded by \i{\{priK A,priK
B\}}. Therefore, after our theorem, if the private keys of \i{A} and
\i{B} are not compromised then the spy cannot have access to \i{NA}
and \i{NB}.

\begin{isa}
\noindent
lemma Guard\_NA: "\Hyp{evs \In nsl; A \Notin bad; B \Notin bad;\\\hs
Says A B (Crypt (pubK B) \pair{Nonce NA, Agent A}) \In set evs}\\\hs
\Imp Guard NA \{priK A,priK B\} (spies evs)"

\noindent
lemma Guard\_NB: "\Hyp{evs \In nsl; A \Notin bad; B \Notin bad;\\\hs
Says B A (Crypt (pubK A) \pair{Nonce NA, Nonce NB, Agent B}) \In set evs}\\\hs
\Imp Guard NB \{priK A,priK B\} (spies evs)"
\end{isa}


We applied our theorem to other well known protocols: the Otway-Rees
sym\-me\-tric-key protocol \cite{otway87} and the Yahalom
sym\-me\-tric-key protocol \cite{abadi90rs}. By doing so, we noticed a
much shorter development time compared with the direct proofs done by
Paulson \cite{paulson98jcs}. Of course, this is not as fast as
automatic tools like TAPS \cite{cohen00csf} but we expect to define
tactics to automate these guardedness proofs.


The Yahalom protocol is interesting since it is a simple case of
dependency between secrets \cite{paulson01jcs}. We recall the informal
definition of the protocol ($S$ denotes the server):

\begin{lst}{}
\item 1. $A\a B: A,Na$
\item 2. $B\a S: B,\{A,Na,Nb\}_{Kb}$
\item 3. $S\a A: \{B,K,Na,Nb\}_{Ka},\{A,K\}_{Kb}$
\item 4. $B\a A: \{A,K\}_{Kb},\{Nb\}_K$
\end{lst}

One can prove that the session key \i{K} is guarded by the keys of
\i{A} and \i{B}, and that the nonce \i{NB} is guarded by the keys of
\i{A} and \i{B} and all the session keys issued by the server:

\begin{isa}
\noindent
lemma Guard\_KAB: "\Hyp{evs \In ya; A \Notin bad; B \Notin bad;\\\hs
Says Server A \pair{Crypt (shrK A) \pair{Agent B, Key K, Nonce NA,
Nonce NB},\\\hs
Crypt (shrK B) \pair{Agent A, Key K}} \In set evs}\\\hs
\Imp Guard K \{shrK A,shrK B\} (spies evs)"

\noindent
lemma Guard\_NB: "\Hyp{evs \In ya; A \Notin bad; B \Notin bad;
Says B Server\\\hs
\pair{Agent B, Crypt (shrK B) \pair{Agent A, Nonce NA, Nonce NB}}
\In set evs}\\\hs
\Imp Guard NB (\{shrK A,shrK B\} $\cup$ keys A B NA NB evs) (spies evs)"

\noindent
keys A B NA NB evs = \{K. Says Server A
\pair{Crypt (shrK A) \pair{Agent B, Key K,\\\hs Nonce NA, Nonce NB},
Crypt (shrK B) \pair{Agent A, Key K}} \In set evs\}
\end{isa}

In our formalization, the server may issue several session keys for
the same request by repeating step 3. This may be avoided by checking
whether step 3 has already been done. Note also that the set of keys
\i{Ks} used here takes care of two different states of the protocol:
if no session key has been issued yet then, indeed, only the keys of
\i{A} and \i{B} are sufficient for protecting \i{NB}.


\section{Guardedness proofs}
\label{sec-proto}

In order to ease guardedness proofs, we have developed other theorems
inspired by the work of Millen and Rue{\ss} \cite{millen00ssp}. The
idea is to reduce the proof for all possible traces to a proof that
all the protocol rules preserve the guardedness condition. This
requires to introduce a more precise formalization of the notion of
protocol.


\begin{definition}[Protocol rule]
A {\em message pattern} is a message with variables of distinct types
for agents, nonces, keys, etc. {\em Event patterns} are defined
similarly. A {\em substitution} \i{s} is an application which
associates an agent to each agent variable, a nonce to each nonce
variable, etc. Replacing each variable of a message pattern \i{X} by
its image in a substitution \i{s} gives a message denoted by \i{apm s
X}. For an event pattern \i{ev}, the substitution is denoted by \i{ap
s ev}.

A {\em rule} \i{R} is a pair made of a set of event patterns \i{fst
R}, the {\em preconditions}, and an event pattern \i{snd R}. The set
of {\em new nonces} of a rule \i{R}, \i{newn R}, is the set of nonce
variables that occur in \i{snd R} and in no precondition.
\end{definition}


Now, a protocol will be seen as a set of rules:

\begin{isa}
\noindent types proto = "(event set * event) set"
\end{isa}

Then, the traces generated by a protocol are inductively defined as
follows:

\begin{isa}
\noindent consts tr :: "proto \arr event list set"\\
inductive "tr p" intros\\
Nil: "[] \In tr p"\\
Fake: "\Hyp{evs \In tr p; X \In synth(analz(spies evs))}\\\hs
\Imp Says Spy B X \# evs \In tr p"\\
Rule: "\Hyp{evs \In tr p; R \In p; ok evs R s} \Imp ap s (snd R)
\# evs \In tr p"
\end{isa}

An \i{s}-instance of a rule \i{R} can be added to a trace \i{evs} if
the \i{s}-instance of every precondition occurs in \i{evs} and the
\i{s}-instance of every new nonce has not been used yet:

\begin{isa}
\noindent ok evs R s = (\All x. x \In fst R \imp ap s x \In set evs)\\\hs
\& (\All n. n \In newn R \imp ap s n \Notin used evs)
\end{isa}


\begin{definition}[Freshness]
We will denote by \i{fresh R' s' n Ks evs} the fact that a nonce \i{n}
is introduced in a trace \i{evs} by an \i{s'}-instance of a rule
\i{R'} in which it is guarded by a set of keys \i{Ks}.
\end{definition}


\begin{theorem}[Guardedness]
Assume that all the rules of the protocol preserve the guardedness of
\i{n} w.r.t. \i{Ks} if the keys of \i{Ks} are safe:

\begin{isa}
\noindent
preserv p n Ks = \All evs R' s' R s. evs \In tr p
\imp fresh R' s' n Ks evs\\\hs
\imp Guard n Ks (spies evs) \imp safe Ks (spies evs)\\\hs
\imp ok evs R s \imp ap s (snd R) \In guard n Ks

\noindent
safe Ks (spies evs) = \All K. K \In Ks \imp K \Notin analz(spies evs)
\end{isa}

Then, if \i{n} is also guarded in the initial knowledge of the spy,
we get that \i{n} is guarded in every possible trace:

\begin{isa}
\noindent theorem preserv\_Guard: "\Hyp{preserv p n Ks; evs \In tr p; 
fresh R' s' n Ks evs;\\\hs
safe Ks (spies evs); Guard n Ks (init Spy)} \Imp Guard n Ks (spies evs)"
\end{isa}
\end{theorem}


\subsection{Examples}

As an example, let us see the case of the Otway-Rees protocol
\cite{otway87}:

\begin{lst}{}
\item 1. $A\a B: \{Na,A,B,\{Na,A,B\}_{Ka}\}$
\item 2. $B\a S: \{Na,A,B,\{Na,A,B\}_{Ka}, \{Na,Nb,A,B\}_{Kb}\}$
\item 3. $S\a B: \{Na,\{Na,K\}_{Ka},\{Nb,K\}_{Kb}\}$
\item 4. $B\a A: \{Na,\{Na,K\}_{Ka}\}$
\end{lst}

The formal definition of the rules are:

\begin{isa}
\noindent OR1 = (\{\},\\\hs
  Says A B \pair{Nonce NA, Agent A, Agent B,\\\hs
  Crypt (shrK A) \pair{Nonce NA, Agent A, Agent B}})

\noindent
OR2 = (\{Says A' B \pair{Nonce NA, Agent A, Agent B, X}\},\\\hs
  Says B Server \pair{Nonce NA, Agent A, Agent B, X,\\\hs
  Crypt (shrK B) \pair{Nonce NA, Nonce NB, Agent A, Agent B}})

\noindent
OR3 = (\{Says B' Server \pair{Nonce NA, Agent A, Agent B,\\\hs
  Crypt (shrK A) \pair{Nonce NA,Agent A, Agent B},\\\hs
  Crypt (shrK B) \pair{Nonce NA, Nonce NB, Agent A, Agent B}}\},\\\hs
  Says Server B \pair{Nonce NA, Crypt (shrK A) \pair{Nonce NA, Key KAB},\\\hs
  Crypt (shrK B) \pair{Nonce NB, Key KAB}})

\noindent
OR4 = (\{Says B Server \pair{Nonce NA, Agent A, Agent B, X,\\\hs
  Crypt (shrK B) \pair{Nonce NA, Nonce NB, Agent A, Agent B}},\\\hs
  Says S B \pair{Nonce NA, Y, Crypt (shrK B) \pair{Nonce NB, Key KAB}}\},\\\hs
  Says B A \pair{Nonce NA, Y})
\end{isa}

Let us prove the preservation property. We only detail the case of
\i{NB}, the case of \i{NA} is similar. Assume that \i{NB} has been
introduced by an instance of \i{OR2} and is guarded by $\{$\i{shrK
B}$\}$.

\begin{lst}{}
\item [\tt OR1:] Assuming that \i{NA'} is a new nonce, we must prove
  that \i{NB} is guarded in \i{P} = \i{\pair{Nonce NA', Agent A',
  Agent B', Crypt (shrK A') \pair{Nonce NA', Agent A',\\ Agent
  B'}}}. Since \i{NA'} is new, it cannot be equal to
  \i{NB}. Guardedness is therefore immediate since \i{NB} does not
  occur in \i{P}.

\item [\tt OR2:] Assuming that \i{NB'} is a new nonce and that \i{NB}
  is guarded in \i{\pair{Nonce NA', Agent A', Agent B', X}}, we must
  prove that \i{NB} is guarded in \i{\pair{Nonce NA', Agent A', Agent
  B', X, Crypt (shrK B') \pair{Nonce NA',\,Nonce NB',\,Agent A',\\
  Agent B'}}}. Again, since \i{NB'} is new, it cannot be equal to
  \i{NB}. Therefore, we are left with the case when \i{NA'} is equal
  to \i{NB}. But, since \i{NB} is guarded in \i{\pair{Nonce NA', Agent
  A', Agent B', X}}, \i{NA'} cannot be equal to \i{NB} and we are
  done.

\item [\tt OR3:] Assuming that \i{NB} is guarded in \i{P} =
  \i{\pair{Nonce NA', Agent A', Agent B', Crypt (shrK A') \pair{Nonce
  NA',Agent A',Agent B'}, Crypt (shrK B') \pair{Nonce\\ NA',Nonce
  NB',Agent A',Agent B'}}}, we must prove that \i{NB} is guarded in
  \i{Q} = \i{\pair{Nonce NA', Crypt (shrK A') \pair{Nonce NA',Key
  KAB}, Crypt\,(shrK B')\,\pair{Nonce\\ NB', Key KAB}}}. Since \i{NB} is
  guarded in \i{P}, \i{NA'} cannot be equal to \i{NB} because \i{NA'}
  is not guarded in \i{P}. Therefore, \i{NB} is guarded in \i{Q}.

\item [\tt OR4:] Assuming that \i{NB} is guarded in \i{\pair{Nonce
  NA',Agent A',Agent B',X, Crypt (shrK B') \pair{Nonce NA', Nonce
  NB', Agent A', Agent B'}}} and in \i{P} = \i{\pair{Nonce NA',Y,Crypt
  (shrK B') \pair{Nonce NB', Key KAB}}}, we must prove that \i{NB}
  is\\ guarded in \i{\pair{Nonce NA',Y}}. Since \i{NB} is guarded in
  \i{P}, it cannot be equal to \i{NA'} and it is guarded in \i{Y}
  also. Therefore, \i{NB} is guarded in \i{Q}.
\end{lst}


With this protocol, the preservation of guardedness does not require
any extra property but, in general, it is necessary to have unicity
lemmas \cite{paulson98jcs,cohen00csf}. Let us see for instance the
Needham-Schroeder-Lowe protocol. Assume that \i{NA} has been
introduced by an instance of \i{NS1} and is guarded by $\{$\i{priK A,
priK B}$\}$. The case of \i{NB} is similar.

\begin{lst}{}
\item [\tt NS1:] Assuming that \i{NA'} is a new nonce, we must prove
  that \i{NA} is guarded in \i{P} = \i{Crypt (pubK B') \pair{Nonce
  NA', Agent A'}}. Since \i{NA'} is new, it cannot be equal to
  \i{NA}. Thus, \i{NA} is guarded in \i{P}.

\item [\tt NS2:] Assuming that \i{NB'} is a new nonce and that \i{NA}
  is guarded in \i{Crypt (pubK B') \pair{Nonce NA', Agent A'}}, we
  must prove that \i{NA} is guarded in \i{P} = \i{Crypt (pubK A')
  \pair{Nonce NA', Nonce NB', Agent B'}}. Again, since \i{NB'} is new,
  it cannot be equal to \i{NA}. So, we are left with the case when
  \i{NA'} = \i{NA}. But then, the trace contains two messages:
  \i{Crypt (pubK B') \pair{Nonce NA, Agent A'}} and \i{Crypt (pubK B)
  \pair{Nonce NA, Agent A}} for \i{NA} has been introduced with
  \i{NS1}. Since \i{NA} is not known to the spy (the trace is assumed
  to be guarded), this cannot happen: we must have \i{A} = \i{A'} (see
  below). Hence, \i{NA} is guarded in \i{P}.

\item [\tt NS3:] Assuming that \i{NA} is guarded in \i{Crypt (pubK B')
  \pair{Nonce NA', Agent A'}} and in \i{Crypt (pubK A') \pair{Nonce
  NA', Nonce NB', Agent B'}}, we must prove that it is guarded in
  \i{P} = \i{Crypt (pubK B') (Nonce NB')}. Assume that \i{NB'} =
  \i{NA}. Then, the trace contains two messages: \i{Crypt (pubK A')
  \pair{Nonce NA', Nonce NA, Agent B'}} and \i{Crypt (pubK B)
  \pair{Nonce NA, Agent A}} for \i{NA} has been introduced with
  \i{NS1}. Since \i{NA} is not known to the spy, this cannot happen
  (see below). Therefore, we must have \i{NB'} $\neq$ \i{NA} and
  \i{NA} is guarded in \i{P}.
\end{lst}

To prove the guardedness, we used the following two unicity lemmas,
which are easily proved:

\begin{isa}
\noindent lemma "\Hyp{evs \In nsl;
Crypt (pubK B) \pair{Nonce NA,\,Agent A} \In parts(spies evs);\\\hs
Crypt (pubK B') \pair{Nonce NA,\,Agent A'} \In parts(spies evs);\\\hs
Nonce NA \Notin analz (spies evs)} \Imp A=A' \& B=B'"

\noindent
lemma "\Hyp{evs \In nsl;
Crypt (pubK B) \pair{Nonce NA,\,Agent A} \In parts(spies evs);\\\hs
Crypt(pubK B') \pair{Nonce NA',\,Nonce NA,\,Agent A'}
\In parts(spies evs)}\\\hs
\Imp Nonce NA \In analz (spies evs)"
\end{isa}

Similar lemmas hold for \i{NB}. They can be seen as two distinct
instances of the same unicity lemma: if two rules that should
introduce new nonces in fact introduce the same nonce in two guarded
messages, and if this nonce is not known to the spy, then the two
messages must be equal.


\subsection{Comparaison with Cortier, Millen and Rue{\ss}' work}

In \cite{millen00ssp}, the proof of unicity lemmas is avoided by
introducing additional information in the definition of protocols, the
{\em spell events}, that indicate the nonces and keys that must be
kept secret and the agents allowed to have access to them, and by
enforcing the disjointness of spell events.

The guardedness proofs often follow the same pattern and use the same
lemmas (that a new nonce is distinct from the nonce for which we want
to prove the guardedness condition, the unicity lemmas, etc.). We
therefore expect to turn this into an Isabelle tactic which would try
to prove the guardedness automatically by using these lemmas.

Cortier, Millen and Rue{\ss} \cite{cortier01csf} define a search
procedure for proving a similar property, the {\em occultness}. But,
since it does not take into account the origin of a secret (the
\i{fresh} predicate in our formalization), it needs to step back into
the protocol rules and the \i{Fake} rule of the spy to get sufficient
information for concluding, and this may not terminate. Furthermore,
they require yet other information in the definitions of protocols,
the {\em state events}, whose usefulness for occultness proofs is not
clear.


\section{Protocols P1 and P2}
\label{sec-def}

As an important and new application, we used our theorem to formally
certify in Isabelle the correctness of some of the protocols proposed
by Asokan, G\"ulc\"u and Karjoth in \cite{asokan98ma} for protecting
the computation results of free-roaming agents. These protocols
tolerate collusion between servers and unfixed itineraries. They
formalize and extend protocols proposed by Yee
\cite{yee98mos}. Application areas of these protocols include
comparison shopping, bidding and network routing
\cite{dicaro98hicss}. Following the authors, we will use the
vocabulary of comparison shopping, hence using the word ``shop'' to
denote a server, and the word ``offer'' to denote the answer of a shop
to an agent's request.

These protocols are not concerned with the security problems raised by
the use of mobile agents: aside from the correctness of the
implementations of servers and mobile agents, that servers indeed
execute the mobile agents code, and that servers and agents are indeed
protected against malicious agents
\cite{yee98mos,buttyan98mos,lee97fsmc}. 

The jump of an agent from a server $S_1$ to a server $S_2$ can be seen
as $S_1$ sending to $S_2$ a message representing the state of the
agent. It therefore fits with our model. The difficulty here is that
the itinerary is not known {\em a priori}: the set of servers to be
visited can be chosen and extended by the agent itself. This is the
case when an agent is programmed to go to some information servers
which may provide it with addresses of shops or other information
servers.

Asokan, G\"ulc\"u and Karjoth propose four protocols called P1, P2, P3
and P4 depending on the properties one would like and the available
infrastructure. P1 and P2 require a public-key infrastructure. In P1,
the author of an offer is not kept secret and the integrity of data is
publicly verifiable while, in P2, the author of an offer can only be
known to the owner of the agent. P3 and P4 do not assume a public-key
infrastructure and use message authentication codes (MAC)
instead. These last two protocols do not ensure non-repudiability. We
formally proved all the properties claimed by the authors for P1 and
P2. We left for future work the proof of P3 and P4.

All the protocols have a common structure: first, the owner of the
agent sends his agent to the first server and, second, each server
receiving the agent sends it to the next server after having added his
own offer, this last step being repeated until the agent comes back to
its owner. The difference between the protocols lie in the way the
messages containing the offers are built.

For describing the state of the agent in P1 and P2, we adopted the
message format \i{\pair{Agent A, Number r, I, L}} where:

\begin{lst}{--}
\item \i{Agent A} is the owner of the agent,
\item \i{Number r} is the agent's request,
\item \i{I} is the list of servers to be visited,
\item \i{L} is the list of offers collected so far.
\end{lst}

We choose to send the request and the list of servers to be visited in
the clear. This is not specified in \cite{asokan98ma} but we may
assume that, in practice, the request is implemented as a public data
and the itinerary as a private data. However, a malicious server
executing the agent could easily know about the itinerary, even though
some encryption is used. So, there is no point in hiding this
information.

On the other hand, we must keep in mind that a malicious server can
get important information from the itinerary stored in the agent and
the way the offers are stored also. Indeed, assume that two servers
$S_1$ and $S_2$ collude. If $S_1$ sends to $S_2$ the itinerary and the
offers of the agent when the agent was at $S_1$, then $S_2$ can try to
guess to whom belong the offers. As suggested in \cite{asokan98ma}, a
way to make this more difficult is to shuffle the offers after each
new offer.

Furthermore, even if a server does not try to alter the offers
collected so far by an agent, it may not give its best offer. Indeed,
if it succeeds to know the offers collected by the agent so far, it
may give an offer which is just a little bit better, but not as good
as it could. This looks like a Vickery auction (the highest bidder
pays the second highest bid) but upside-down: the best server offers
the second lowest price \cite{yee98mos}.

Then, we formalized P1 and P2 as follows:

\begin{isa}
\noindent consts p :: "event list set"\\
inductive p intros\\
Nil: "[] \In p"\\
Fake: "\Hyp{evs \In p; X \In synth(analz(spies evs))}\\\hs
  \Imp Says Spy B X \# evs \In p"\\
Req: "\Hyp{evs \In p; Nonce n \Notin used evs; I \In agl}\\\hs
  \Imp Says A B (reqm A r n I B) \# evs \In p"\\
Prop: "\Hyp{evs \In p; I \In agl; J \In agl;
Says A' B \pair{Agent A,Number r,I,L}\\\hs \In set evs;
isin (Agent C, app (J, del (Agent B, I))); Nonce ofr\\\hs \Notin used evs}
\Imp Says B C (prom B ofr A r P I L J C) \# evs \In p"\\
\end{isa}

\i{Nil} and \i{Fake} are the usual rules for the empty trace and the
spy respectively. \i{Req} corresponds to the first step, the sending
of the agent by its owner. We use an unguessable message, \i{Nonce n},
to identify the session. \i{Prop} corresponds to the second step, the
addition of an offer by a server. The offer is represented by an
unguessable nonce, \i{Nonce ofr}. \i{agl} is the subset of messages
representing lists of agents. \i{isin (Agent C, app (J, del (Agent B,
I)))} means that the next server \i{C} is picked among the agents of
\i{I}, except the first occurrence of \i{B}, or among a new list of
agents \i{J}. Finally, \i{reqm} and \i{prom} are the request message
and the proposition message respectively. They are specific to each
protocol.

The properties claimed to hold for these two protocols are the following:

\begin{lst}{--}
\item {\em Data confidentiality}:
only the owner of the agent can extract the offers.

\item {\em Non-repudiability}:
a shop cannot repudiate an offer once it has been received by the
owner of the agent.

\item {\em Forward privacy}: (for P2)
none of the identities of the shops can be extracted.

\item {\em Strong forward integrity}:
except the last one, offers cannot be modified.

\item {\em Publicly verifiable forward integrity}: (for P1 only)
anyone can verify the integrity of a list of offers.

\item {\em Insertion resilience}:
no offer can be inserted between two previous offers.

\item {\em Truncation resilience}:
the list of collected offers can be truncated only at an offer whose
author colludes with the attacker.
\end{lst}

Note that forward privacy only means that no one can extract the
identity of the shops by looking only at the list of offers. This does
not mean that the identity of shops cannot be inferred by other means
as described above.

The idea proposed by Asokan, G\"ulc\"u and Karjoth for ensuring
insertion resilience, truncation resilience and a stronger form of
forward integrity than the one proposed by Yee \cite{yee98mos} is to
add unforgeable dependencies between offers, creating what they call a
{\em chaining relation}:

\begin{lst}{--}
\item The agent starts with a hash of the message made of the identity
of the first server to be visited \i{B}, salted with some random
number \i{k}: \i{Hash \pair{Agent B, Nonce k}}

\item Then, within its offer, signed with its private key, each shop
must add a hash of the message made of the previous offer \i{M}
together with the next server \i{C} to be visited: \i{Hash \pair{M,
Agent C}}.
\end{lst}

The complete definition of \i{chain B ofr A L C} is as follows for P1:

\begin{isa}
\noindent
sign B \pair{Crypt (pubK A) (Nonce ofr), Hash \pair{head L,Agent C}}
\end{isa}

and as follows for P2:

\begin{isa}
\noindent
\pair{Crypt (pubK A) (sign B (Nonce ofr)), Hash \pair{head L,Agent C}}
\end{isa}

\noindent
where \i{head L} denotes the previous offer collected by the agent and
\i{sign} is the signature function defined by \i{sign B X =
\pair{Agent B, X, Crypt (priK B) (Hash X)}}. The start of the chaining
relation, \i{anchor}, is defined as a particular case of \i{chain} by
\i{anchor A n B = chain A n A (cons nil nil) B}.

Hence, if someone wants to modify or insert an offer, for the chaining
relation to be preserved, he must be able to modify as well all the
following offers, which is made {\em a priori} impossible by asking
the shops to sign their offers with their private keys. The two
remaining possible attacks are the deletion of the offers between two
offers made by two colluding shops (or the same shop if the agent goes
twice to the same malicious shop) or the deletion of all the offers
(denial-of-service attack), and the addition of fake offers. So, a
shop cannot even modify its own offer unless it colludes with another
shop visited later (which may be the same) but, in this case, it must
delete all the intermediate offers. For limiting this difficult
deletion/truncation problem, Asokan, G\"ulc\"u and Karjoth suggest a
few solutions like adding several next shops instead of just one.

In our formalization, we do not need to include the salt random
numbers $r_i$ used in \cite{asokan98ma} since the spy is not able to
infer the content of encrypted messages if he does not know the
inverse of the keys used for encrypting them (\i{analz} function).

We can know present the formal definitions of the requests and
propositions:

\begin{isa}
\noindent
reqm A r n I B = \pair{Agent A, Number r,
cons (Agent A) (cons (Agent B) I),\\\hs
cons (anchor A n B) nil}

\noindent
prom B ofr A r I L J C = \pair{Agent A,Number r,
app (J, del (Agent B, I)),\\\hs
cons (chain B ofr A L C) L}
\end{isa}

In the request, \i{A} and \i{B} (the first server to be visited) are
added to the itinerary \i{I}. In the proposition, \i{B} is deleted
from \i{I} and a new list of servers \i{J} is added.


\section{Correctness of P1 and P2}
\label{sec-proof}

For proving the strong forward integrity, the insertion resilience and
the truncation resilience, we need to define what is a {\em valid}
chaining relation:

\begin{isa}
\noindent inductive "valid A n B" intros\\
Req:~"cons (anchor A n B) nil \In valid A n B"\\
Prop:~"L \In valid A n B\\\hs
\Imp cons (chain (next\_shop(head L)) ofr A L C) L \In valid A n B"
\end{isa}

And to formalize the corresponding attacks, we use the following
functions:

\begin{lst}{--}
\item \i{ith(L,i)} is the \i{i+1}-th element of \i{L}.
\item \i{repl(L,i,M)} is the list \i{L} with its \i{i+1}-th
element replaced by \i{M}.
\item \i{ins(L,i,M)} is the list \i{L} with \i{M} inserted
before the \i{i+1}-th element of \i{L}.
\item \i{trunc(L,i)} truncates the \i{i} first elements of \i{L}.
\end{lst}

The three properties are then easily proved by induction on \i{valid}:

\begin{isa}
\noindent lemma strong\_forward\_integrity:
"\Hyp{L \In valid A n B; Suc i < len L;\\\hs
repl(L,Suc i,M) \In valid A n B} \Imp M = ith(L,Suc i)"

\noindent
lemma insertion\_resilience:
"\Hyp{L \In valid A n B; Suc i < len L}\\\hs
\Imp ins(L,Suc i,M) \Notin valid A n B"

\noindent
lemma truncation\_resilience:
"\Hyp{L \In valid A n B; Suc i < len L;\\\hs
cons M (trunc(L,Suc i)) \In valid A n B} \Imp shop M = shop (ith(L,i))"
\end{isa}

For the data confidentiality, we first prove that both a request \i{n}
and an offer \i{ofr} are guarded by the private key of the agent's
owner. Then:

\begin{isa}
\noindent lemma req\_notin\_spies: "\Hyp{evs \In p1;
req A r n I B \In set evs; A \Notin bad}\\\hs
\Imp Nonce n \Notin analz (spies evs)"

\noindent
lemma pro\_notin\_spies: "\Hyp{evs \In p1;
pro B ofr A r I L J C \In set evs;\\\hs A \Notin bad;
B \Notin bad} \Imp Nonce ofr \Notin analz (spies evs)"
\end{isa}

We also proved that requests and offers are not known by other agents
(and not only by the spy as it is commonly done). Although this is not
very complicated, this requires to extend the Isabelle libraries.

For the non-repudiability, we proved that the signature scheme is
secure:

\begin{isa}
\noindent
lemma "\Hyp{evs \In p1; A \Notin bad; sign A X \In parts(spies evs)}\\\hs
\Imp \Ex B Y. Says A B Y \In set evs \& sign A X \In parts {Y}"
\end{isa}

We would like to point out that, although it was the first time we
used Isabelle, thanks to the way we formalized P1 and the power of the
Isabelle tactics (although, sometimes, we would like them to be less
sensitive to some syntactical aspects), once P1 has been proved, it
took us only a few minutes to formalize and prove P2 by essentially
changing the definition of \i{chain}. It is easy to experiment with
changes in message formats.


\section{Conclusion and future work}
\label{sec-conclu}

Approaches based on proof assistants like Paulson's inductive approach
\cite{paulson98jcs} in Isabelle \cite{nipkow02lncs} are known to
require much effort, typically several days or weeks, for certifying a
protocol. Establishing general protocol-independent theorems like the
ones we presented in this paper helps to reduce this development time
and also to get a better understanding of protocol problems. To go
further, automated theorem provers could be used or tactics developed
for automatically proving the conditions of these theorems. In the
case of our guardedness condition, it is clear from the examples we
give that the proofs follow similar patterns. This is why we expect to
define Isabelle tactics for doing that and also for proving the
unicity lemmas automatically.

Finally, we did not take into account in our formalizations what some
authors call the ``oops'' rule \cite{paulson01jcs}, that is, the fact
that, for some protocols like Yahalom, a particular session is not
affected by the compromise of other session keys. But, instead of
adding the oops rule and doing the proof again, one may look for
conditions under which a secrecy property without oops implies the
same property with oops. This would provide another important and
useful protocol-independent result.\\

{\bf Acknowledgments.} This work has been done during my stay at
Cambridge (UK) in 2002 thanks to a grant from the INRIA (French
National Research Institute in Computer Science and Automatics) and
the {\sc epsrc} grant GR\slash R01156\slash R01 \emph{Verifying
Electronic Commerce Protocols}. We also want to thank Larry Paulson
and Glynn Winskel for their comments on this paper.



\end{document}